\documentclass[lettersize,journal]{IEEEtran}

\usepackage{amsmath,amsfonts}
\usepackage{algorithmic}
\usepackage{algorithm}
\usepackage{array}
\usepackage[caption=false,font=normalsize,labelfont=sf,textfont=sf]{subfig}
\usepackage{textcomp}
\usepackage{stfloats}
\usepackage{url}
\usepackage{verbatim}
\usepackage{graphicx}
\usepackage{cite}
\usepackage{multirow}
\usepackage{booktabs}
\usepackage{xcolor}
\usepackage{enumitem} 

\begin{document}

\title{AI/ML Life‑Cycle Management for Interoperable AI‑Native RAN}

\author{Chu-Hsiang Huang,~\IEEEmembership{Member,~IEEE},~Yuan-Chih Fan Chiang,~\IEEEmembership{Student Member,~IEEE}, ~Chao-Kai~Wen,~\IEEEmembership{Fellow,~IEEE},~and~Geoffrey~Ye~Li,~\IEEEmembership{Fellow,~IEEE}

\thanks{{C.-H.~Huang and Y.-C. Fan Chiang} are with the Department of Electrical Engineering, National Taiwan University, Taipei 10617, Taiwan, Email: {\rm \{chuhsianh, r14942059\}@ntu.edu.tw}.}
\thanks{{C.-K.~Wen} is with the Institute of Communications Engineering, National Sun Yat-sen University, Kaohsiung 80424, Taiwan, Email: {\rm chaokai.wen@mail.nsysu.edu.tw}.}
\thanks{{G.~Y.~Li} is with the Department of Electrical and Electronic Engineering, Imperial College London, SW7 2AZ London, U.K., Email: {\rm geoffrey.li@imperial.ac.uk}.}

\thanks{This article has been accepted for publication in IEEE Communications Standards Magazine.

{\copyright~2026 IEEE. Personal use of this material is permitted.
Permission from IEEE must be obtained for all other uses, in any current or
future media, including reprinting/republishing this material for advertising
or promotional purposes, creating new collective works, for resale or
redistribution to servers or lists, or reuse of any copyrighted component of
this work in other works.}}

}

\markboth{IEEE Communications Standards Magazine}%
{Shell \MakeLowercase{\textit{et al.}}: A Sample Article Using IEEEtran.cls for IEEE Journals}

%% Remember, if you use this you must call \IEEEpubidadjcol in the second
%% column for its text to clear the IEEEpubid mark.

%TC:ignore
\maketitle

\begin{abstract}
Artificial intelligence (AI) and machine learning (ML) are rapidly becoming integral to the 5G Radio Access Network (RAN), enabling beam management, channel state information (CSI) feedback, positioning, and mobility prediction. However, without a standardized life-cycle management (LCM) framework, challenges such as model drift, vendor lock-in, and limited transparency hinder large-scale deployment. 3GPP Releases 17--20 have progressively introduced AI/ML management and air-interface support, covering model training, validation, deployment, inference, data collection, performance monitoring, applicability assessment, and feature-specific control. Release 20 further extends these capabilities to two-sided CSI compression and inter-vendor model operation. This article reviews the resulting five-block LCM architecture, KPI-driven monitoring mechanisms, and inter-vendor collaboration schemes. We further propose an enhanced LCM framework with detailed interactions across functional blocks and an integrated procedure for reference-model and vendor-model development in two-sided operation, and identify open challenges in resource-efficient monitoring, environment drift detection, intelligent decision-making, and flexible model training. These developments provide a foundation for AI-native transceivers in 6G.
\end{abstract}
%TC:endignore

%TC:ignore
\begin{figure*}[!t]
 \centering
 \includegraphics[width=6.5in]{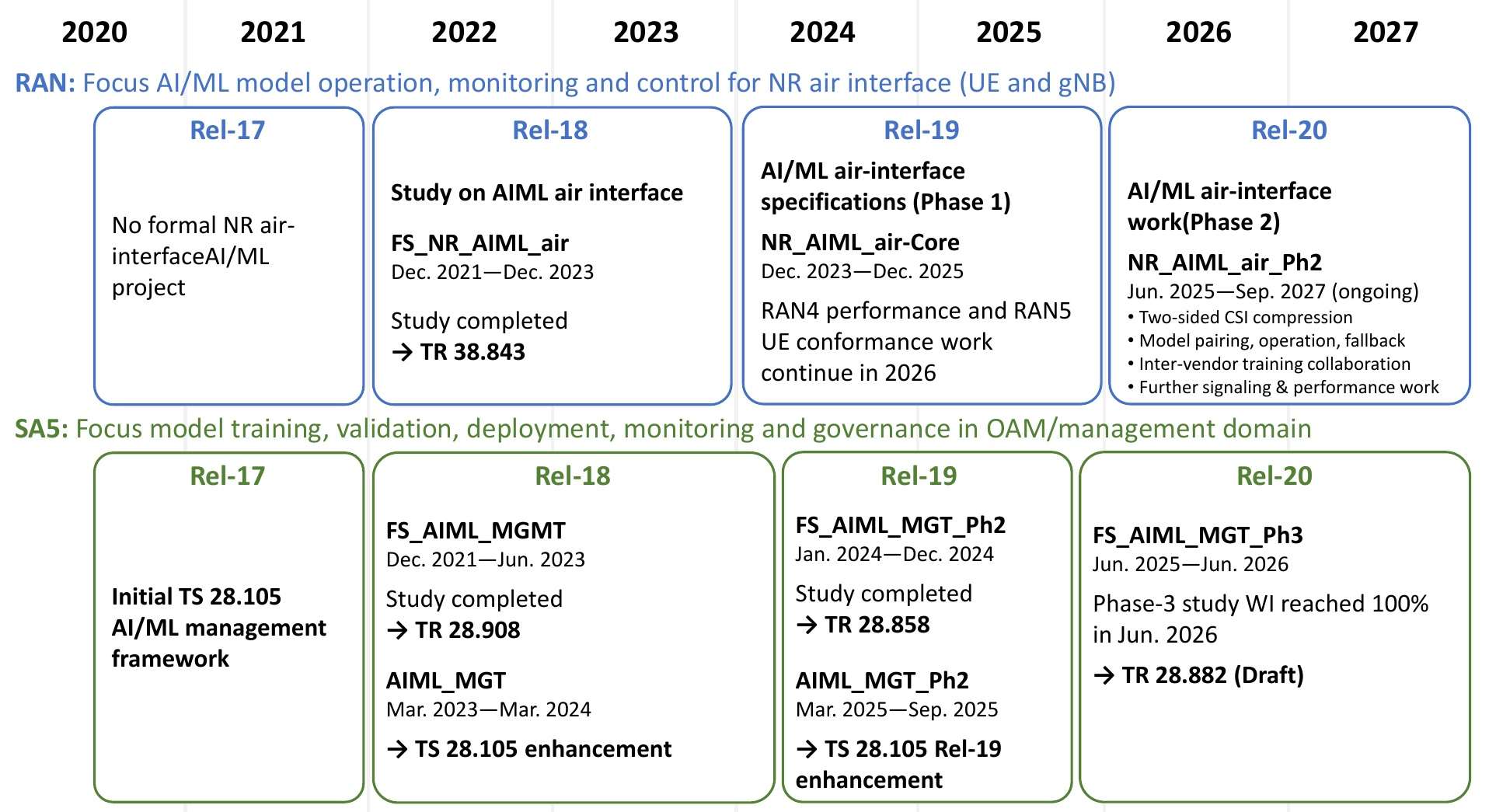}
 \caption{3GPP AI/ML LCM Timeline.}
  \label{fig:lcm_timeline} 
  \vspace{-0.3cm}
\end{figure*}
%TC:endignore

\section{Introduction}

\IEEEPARstart{A}{rtificial} intelligence (AI) and machine learning (ML) have demonstrated significant potential for enhancing radio access network (RAN) performance, particularly in nonlinear and analytically complex tasks such as beam management \cite{Xue-VTMag24}, channel state information (CSI) feedback \cite{Guo-22TCOM,Korpi-26WCOM}, and mobility prediction \cite{TR38744}. As AI/ML models become increasingly integrated into communication systems, their life cycles, including training, deployment, inference, monitoring, and replacement, must be properly managed to maintain reliable operation under diverse and evolving conditions. Large-scale deployment further introduces challenges related to model drift, version consistency, performance degradation, and cross-vendor compatibility. These challenges motivate a standardized life-cycle management (LCM) framework that can sustain model performance and support adaptation across heterogeneous and multi-vendor networks \cite{Lin2023AI3GPP,Lin2024AI3GPP}.
%positioning \cite{Foliadis-24TMLCN}, 

On the management side, \textbf{Release 17 (Rel-17)} marked the initial formalization of generic AI/ML management in the 3rd Generation Partnership Program (3GPP) Service and System Aspects Working Group 5 (SA5). TS~28.105 introduced management capabilities for ML model training and re-training, providing an initial integration of AI/ML operations into existing operations, administration, and maintenance (OAM) processes. These capabilities were subsequently extended in Rel-18 into a broader management framework covering additional phases of the AI/ML life cycle.

On the RAN side, \textbf{Release 18 (Rel-18)} established a distinct standardization track for AI/ML-enabled New Radio (NR) air-interface operation. RAN1 completed the Study Item (SI) \texttt{FS\_NR\_AIML\_air}, with its findings documented in TR~38.843~\cite{TR38843}. The study examined CSI feedback enhancement, beam management, and positioning, together with their performance, complexity, interoperability, and potential specification impacts.  
In parallel, SA5 developed an OAM-oriented AI/ML operational framework in TR~28.908~\cite{TR28908}, covering model training, validation, testing, inference emulation, deployment, and inference. The corresponding normative management capabilities were specified mainly in Rel-18 TS~28.105. These longer-timescale OAM functions support the overall model life cycle but should be distinguished from the latency-sensitive monitoring and control procedures required for AI/ML operation over the NR air interface.

\textbf{Release 19 (Rel-19)} converted selected outcomes of the Rel-18 RAN study into normative air-interface support under the \texttt{NR\_AIML\_air} Work Item (WI). The work addressed AI/ML-assisted CSI prediction, beam management, and positioning, together with feature-specific mechanisms for data collection, applicability assessment, performance monitoring, and related Radio Resource Control (RRC) configuration.  

The core Rel-19 air-interface work was completed in 2025, while associated performance evaluation and User Equipment (UE) conformance activities continue through 2026.

\textbf{Release 20 (Rel-20)} extends the RAN work toward two-sided AI/ML model operation. In June 2025, 3GPP initiated WI \texttt{NR\_AIML\_air\_Ph2}, which remains ongoing and is scheduled to continue through 2027. One of its main topics is two-sided CSI compression, in which an encoder at the UE and a decoder at the gNB jointly support CSI feedback. In the Case-0 setting, the compression operates on spatial- and frequency-domain CSI without exploiting temporal information. The ongoing work considers model pairing and operation, inter-vendor training collaboration, signaling, monitoring, fallback, and performance evaluation. 

In addition to the continued evolution of 5G-Advanced, Rel-20 initiates the 3GPP study phase for 6G Radio \cite{2025066GSID}, with the first normative 6G specifications planned for Rel-21. AI/ML-enabled air-interface techniques form an important part of this investigation, where effective LCM and model interoperability will be necessary for AI-native transceiver functions, particularly when cooperating models are distributed across the UE and network sides. The standardized air-interface LCM and interoperability mechanisms developed in 3GPP can provide underlying observation and control capabilities for broader AI-RAN functions, such as Telemetry, Monitoring, Observability \& Analytics and the Multi-Agent Manager described in \cite{AIRAN}. These specifications therefore provide an important foundation for implementing the broader AI-native RAN architecture envisioned by the AI-RAN Alliance.

As illustrated in Fig.~\ref{fig:lcm_timeline}, 3GPP AI/ML life-cycle standardization has evolved along two related but distinct tracks. The SA5 track provides longer-timescale OAM capabilities for model training, validation, testing, deployment, inference management, and life-cycle governance. In contrast, the RAN track addresses time-sensitive model operation, monitoring, applicability assessment, signaling, and interoperability over the NR air interface. Rel-18 established the principal frameworks for these two tracks, Rel-19 introduced normative enhancements, and Rel-20 continues the development of two-sided AI/ML air-interface operation.

In contrast to \cite{Lin2023AI3GPP,Lin2024AI3GPP}, which mainly review the progress of AI/ML standardization in 3GPP Rel-18, this article is from a RAN-centric view of operational LCM beyond Rel-18. SA5 OAM mechanisms are introduced only where they provide necessary longer-timescale life-cycle support.
Section~\ref{sec:lcm} reviews the RAN-oriented LCM framework and its functional components.
Section~\ref{sec:interop} analyzes interoperability schemes for two-sided models and their trade-offs.
Section~\ref{sec:keytech} identifies key technical challenges and future research directions for interoperable AI-native RAN systems.

%TC:ignore
\begin{figure*}[!t]
 \centering
 \includegraphics[width=6.0in]
 {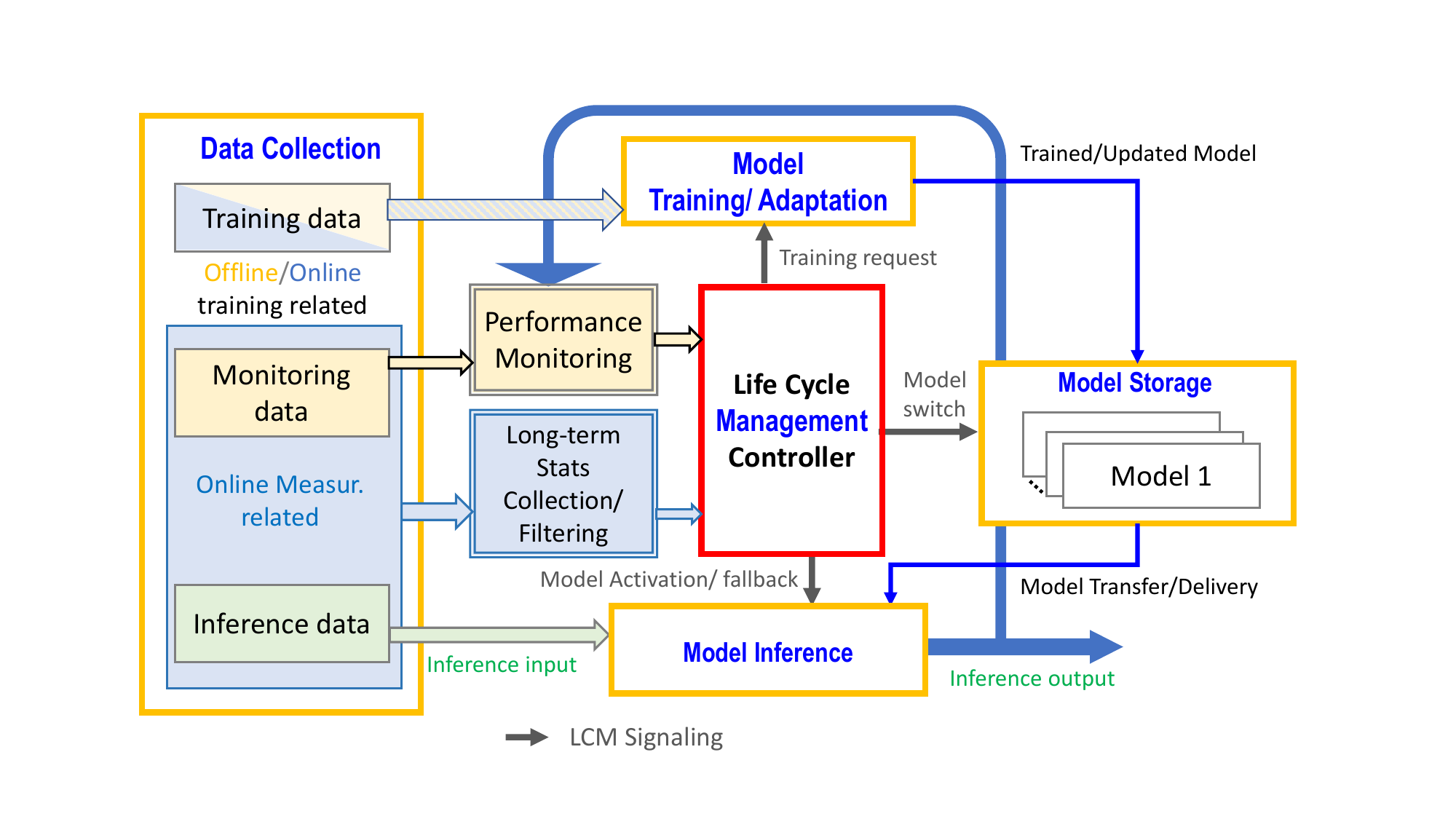}
 \caption{RAN-oriented AI/ML operational LCM and its interaction with longer-timescale OAM life-cycle support.}
 %\caption{Detailed interaction of AI/ML LCM sub-functions in 3GPP.}
  \label{fig:lcm_framework_detailed} 
\end{figure*}
%TC:endignore
 
\section{{RAN-Centric} LCM Architecture} \label{sec:lcm}
% \section{LCM Architecture {\red in RAN}} 
{In this section, we propose the RAN-side operational LCM framework based on the 3GPP discussions in TR~38.843 {and our analysis of the LCM functional blocks}, with particular emphasis on performance monitoring and model control over the NR air interface. We then use CSI prediction and compression as a case study of how the functional blocks in the LCM framework interact with each other in one of the 3GPP AI/ML use cases.}

\subsection{Functional Framework}\label{subsec:lcm_framework} 
{We propose an LCM functional framework based on the five logical functions for the NR air interface defined in TR~38.843 \cite{TR38843}, as illustrated in Fig.~\ref{fig:lcm_framework_detailed}. The 3GPP-defined blocks are highlighted by orange and red boxes and connected by black arrows. Double-lined boxes represent additional functional elements introduced in this article, and their associated arrows indicate the interactions required to support the derivation of appropriate control actions.}

In the 3GPP functional framework, the \textbf{Management} function logically coordinates the operation of the \textbf{Data-Collection}, \textbf{Model-Training/Adaptation}, \textbf{Model-Storage}, and \textbf{Inference} blocks, thereby closing the performance-governance loop. 

The Management function in \cite{TR38843} is a logical function and does not prescribe a single physical entity or a common operating timescale for all management actions. In this article, we treat a fast RAN-local controller and a slower OAM life-cycle manager differently. The RAN-local controller handles monitoring configuration, applicability assessment, activation or deactivation, switching among available models, and temporary fallback. The OAM life-cycle manager supports training, validation, model loading, repository management, and retraining.

{Based on the five logical functions in \cite{TR38843}, the proposed framework further integrates the following functional elements:}
\begin{itemize}
\item \textbf{Monitoring}  schedules and collects observations used for evaluation of AI/ML model operation and performance.
\item \textbf{Key Performance Index (KPI) derivation}  transforms raw observations into measurable indicators of accuracy, robustness, and trustworthiness of the AI/ML models.
\item \textbf{Environment detection} compares the current data distributions against those observed during model training.

\item {\textbf{Adaptation} supports fast and lightweight RAN-local model weight/parameter adjustment as a complementary mechanism to model training/retraining which is usually performed offline within longer timescales.}

\item {\textbf{Decision logic} integrates multiple inputs and selects an appropriate RAN-local action, such as maintaining the current model, switching among available models, or temporarily falling back to a non-AI procedure. Persistent degradation may be reported to the slower OAM loop for model replacement or retraining.}
\end{itemize}

The proposed LCM framework in Fig.~\ref{fig:lcm_framework_detailed} integrates the associated LCM functionalities and define the proper interactions between the blocks to fulfill these functional requirements.
Within the RAN-local loop, inference and monitoring data are collected to evaluate the operation of the active model. The inference block uses the received signals to produce real-time outputs, while the performance-monitoring function derives short-term KPIs from the inference outputs and corresponding reference signal measurements. The management part of LCM takes all the information, including the one from the measurements and the model output, to make the model control decisions. The controller then coordinates the storage, training/adaptation, and inference blocks to execute the control decisions via internal or external signaling. Aggregated monitoring data and environment statistics may also be forwarded to the slower OAM loop. In the OAM loop, these data can support model training, validation, updating, and repository management.

{The RAN-based controller evaluates whether the currently active model remains suitable. When short-term performance falls below a predetermined criterion, it may maintain the current model, deactivate the AI/ML function, switch among already available models, or temporarily fall back to a non-AI procedure. The RAN-based LCM can be either functionality- or model-ID-based, and accompanying signaling frameworks are included in \cite{TR38843}. Persistent degradation or significant changes in input statistics may be reported to the OAM life-cycle manager, which can initiate model validation, replacement, or retraining on a longer timescale.} 

\subsection{Performance Monitoring {in RAN}}\label{subsec:perf_kpis}

Performance monitoring provides the measurement-based information needed for LCM decisions in the RAN. Because radio conditions and inference quality can change on relatively short timescales, RAN-side monitoring is needed to evaluate whether an active AI/ML function remains applicable and performs as intended. Compared with longer-timescale OAM performance management, RAN-based performance monitoring enables the wireless communication system to react to the physical environment change within much smaller timescales. 
The performance-monitoring requirement study for AI/ML-enabled NR air-interface functions was captured in \cite{TR38843} and Rel-19 specifications for performance monitoring procedures and requirements are based on the study in \cite{TR38843}.

To support effective monitoring, appropriate monitoring resources must be configured for the corresponding inference events. These measurements enable ground-truth comparison and evaluation of model performance. For instance, in beam prediction tasks, the UE measures received signal strength or CSI on the beams predicted by the model. 
The gNB configures the monitoring resources and associates them with the corresponding prediction or inference reports so that the resulting measurements can be used for KPI computation.

{TR~38.843 \cite{TR38843} considers multiple evaluation metrics for different AI/ML use cases.} Representative examples include: 
\begin{itemize}
    \item \textbf{Beam prediction accuracy:} As defined in \cite{TR38843}, this metric counts how often, among the $N$ most recent monitoring intervals, at least one of the $M$ best measured beams appears within the top-$K$ predicted beams.
    \item \textbf{CSI prediction quality:} The squared generalized cosine similarity (SGCS) metric quantifies the alignment between predicted precoder $\tilde{w}$ and true precoder $w$ as follows:
    \begin{equation} 
        SGCS = \frac{|\tilde{w}^{H} w|^2}{\|\tilde{w}\|^2 \|w\|^2}.
    \end{equation}
    \item {\textbf{Positioning accuracy:} According to \cite{TR38843}, the 90\% cumulative distribution function (CDF) percentiles of horizontal accuracy is taken as the performance metrics for positioning accuracy evaluation.}
\end{itemize}
Table~\ref{tab:kpi_matrix} summarizes these and other evaluation metrics across common AI/ML use cases. 
Depending on the use case, performance metrics may be computed at the UE, the gNB, or jointly through UE-assisted reporting. The resulting measurements support RAN-local decisions, such as continued operation, temporary deactivation, switching among available models, or fallback.

%TC:ignore
\begin{table}[!t]
\centering
\caption{Representative Evaluation Metrics for AI/ML RAN Use Cases}
\label{tab:kpi_matrix}
\renewcommand{\arraystretch}{1.2}
\begin{tabular}{|p{2.0cm}|p{3.3cm}|p{2.3cm}|}
\hline
\textbf{Use-case} & \textbf{Evaluation Metrics} & \textbf{Evaluation Point} \\ \hline
Beam management & Top-$K$ beam prediction accuracy & gNB or UE assisted \\ \hline
CSI compression & SGCS & gNB or UE-assisted  \\ \hline
Positioning & Position-error CDF & gNB \\ \hline
\end{tabular}
\end{table}
 %TC:endignore

\subsection{Case Study: CSI Prediction and Compression}\label{subsec:case_study}
\begin{figure*}[!t]
  \centering
  \includegraphics[width=0.75\linewidth]{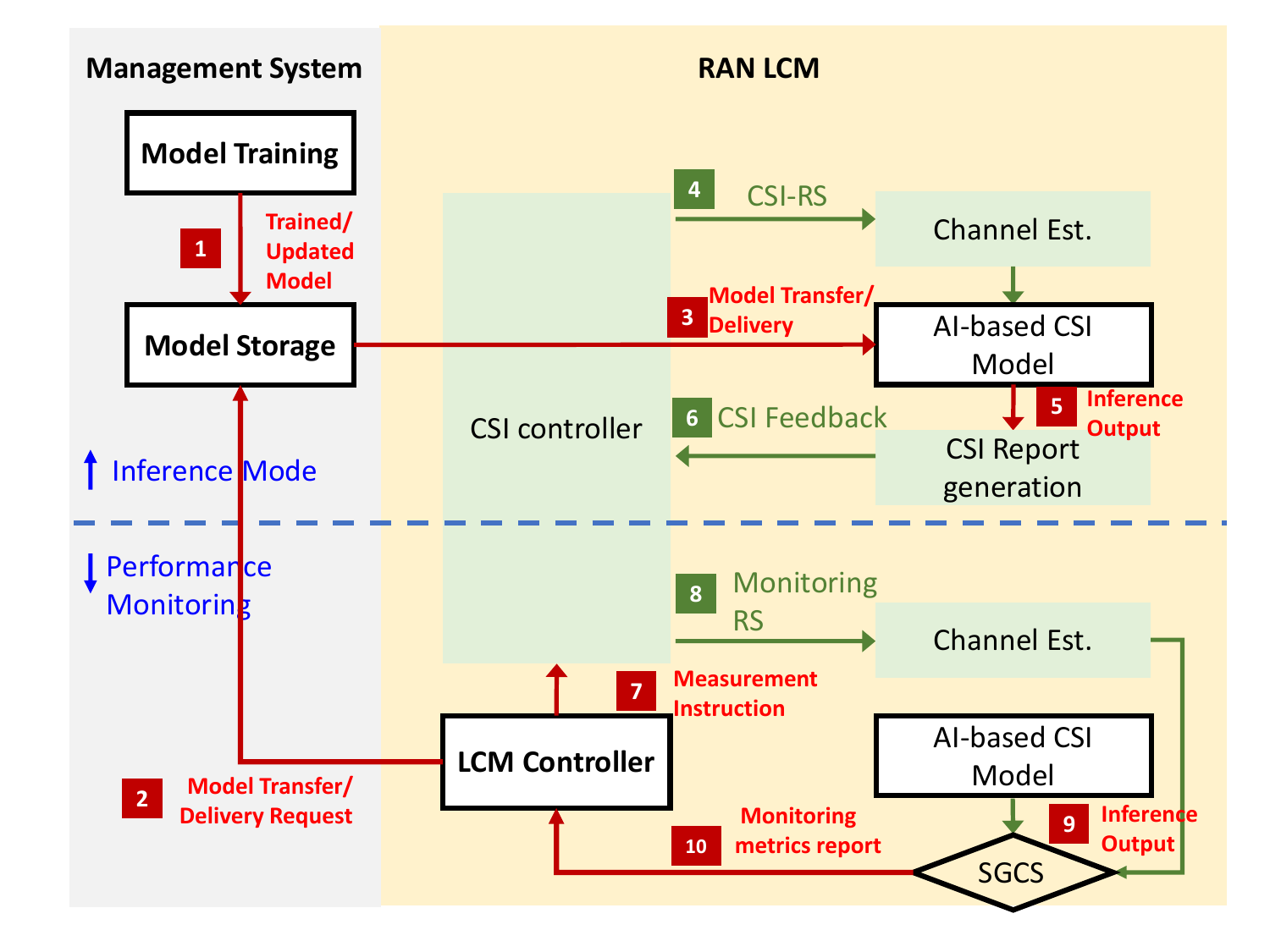}
  \caption{One-sided CSI prediction mapped onto the 3GPP AI/ML LCM framework. Red arrows indicate LCM control paths; green arrows represent legacy CSI reporting.}
  \label{fig:csi_flow_lcm}
\end{figure*}

In 5G networks, CSI reports are typically generated every 40--160~ms. 
This periodicity introduces \emph{channel aging}, particularly for medium- and high-mobility UEs, whose instantaneous channel states may have drifted significantly since the last report, leading to system performance degradation.  
While increasing the reporting frequency could mitigate this degradation, it would also introduce substantial uplink overhead.
CSI reporting overhead can be reduced through CSI prediction or CSI compression. In CSI prediction, the UE or gNB predicts the CSI corresponding to a future transmission time. In CSI compression, the UE encodes the estimated CSI into a compact latent representation, which is reported to the gNB for reconstruction. 3GPP studied both approaches in Rel-18. CSI prediction was subsequently included in the Rel-19 normative work, while two-sided CSI compression is being addressed in the ongoing Rel-20 work. Section~\ref{sec:interop} discusses the interoperability issues associated with two-sided CSI compression, whereas this section focuses on their LCM and performance-monitoring aspects.
The LCM and performance-monitoring functions introduced earlier apply to these use cases. {Fig.~\ref{fig:csi_flow_lcm} illustrates an example of how these functions supervise a UE-side CSI model.}

%TC:ignore
%TC:endignore

\paragraph*{Model Provisioning and RAN-Side Operation} 
{Model training may be performed by an external training platform or management system, and validated models may subsequently be loaded into the relevant RAN entity.}
During inference, the UE collects Channel State Information Reference Signal (CSI-RS) measurements, performs local prediction and/or compression, and reports the PMI or latent message with optional CQI and RI extensions to the gNB.\footnote{PMI: Precoding Matrix Indicator; CQI: Channel Quality Indicator; RI: Rank Indicator.}  
{The RAN-local controller uses the configured performance-monitoring results to determine whether the active model remains suitable or whether temporary deactivation, model switching, or fallback is required.}

\paragraph*{Performance Monitoring} 
For performance monitoring, the gNB shall schedule reference signals as monitoring resources, including CSI-RS, Demodulation Reference Signal (DM-RS) or Sounding Reference Signal (SRS) resources, to obtain or infer the ground-truth CSI.
Rel-19 introduced UE-assisted performance monitoring for CSI prediction, including SGCS-based reporting configured through the CSI reporting framework. The detailed reporting content and associated reference resources depend on the configured monitoring procedure. CSI compression monitoring schemes are under discussion in Rel-20.  
\begin{figure}[h]
  \centering
  \includegraphics[width=0.8\linewidth]{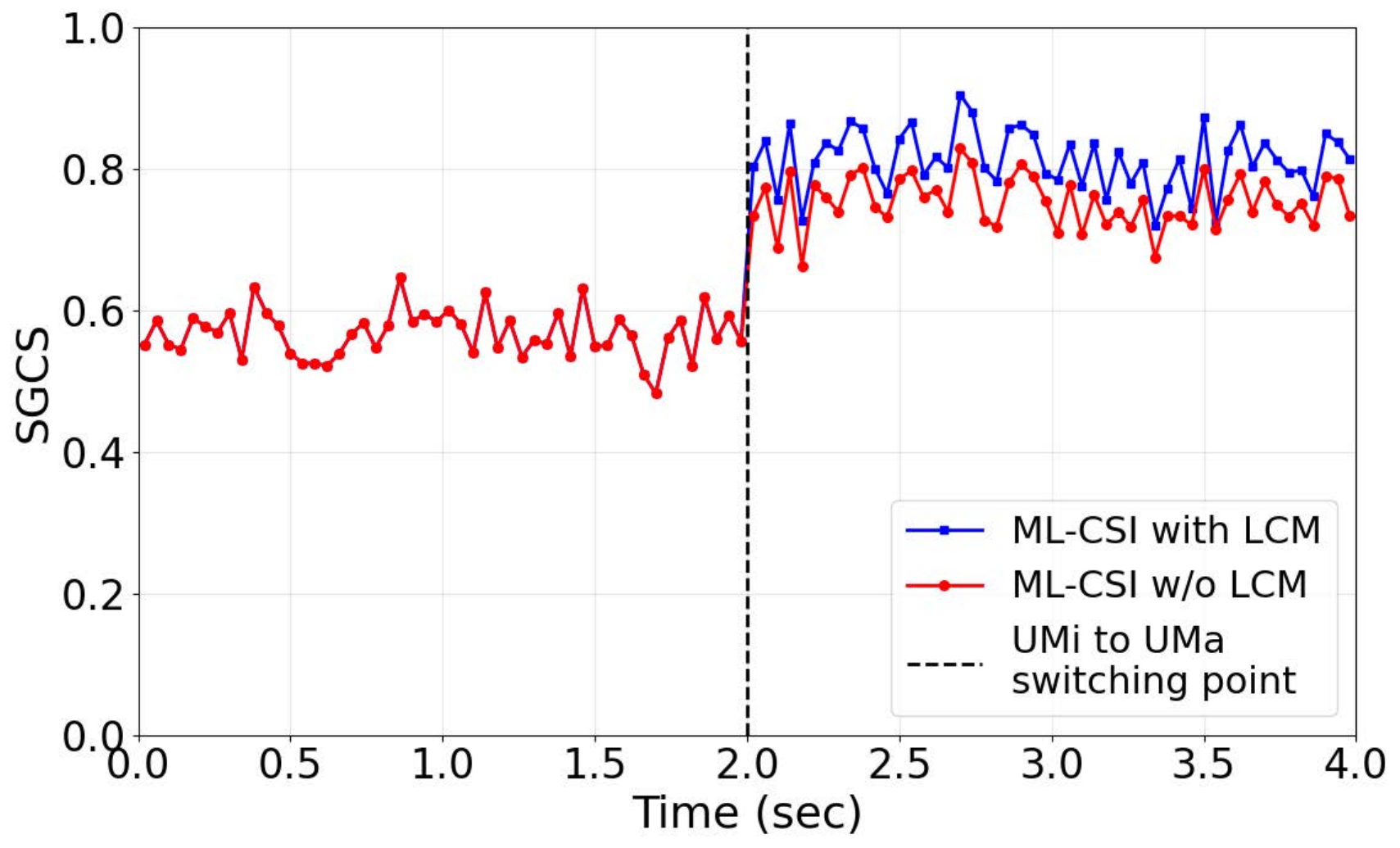}
  \caption{Case study of LCM for ML-CSI use case.}
  \label{fig:csi_sim}
  \vspace{-0.3cm}
\end{figure}

\paragraph*{LCM Simulations} 
{We illustrate the value of LCM through a CSI compression case study evaluated using SGCS, as introduced in Sec.~\ref{subsec:perf_kpis}. The ML-based CSI compression scheme in \cite{Guo-25TWireless} is evaluated under an environment change from Urban Micro (UMi) to Urban Macro (UMa), both of which are representative 5G scenarios defined in \cite{TR38901}. Two models are trained using the UMi and UMa datasets, respectively. As shown in Fig.~\ref{fig:csi_sim}, without an LCM controller, the system continues using the UMi-based model after the environment change, whereas the LCM controller switches to the UMa-based model, resulting in an approximately 10\% SGCS improvement. Notably, the SGCS may increase after the transition to UMa even when the mismatched UMi-based model is retained because the UMa channel is less complex. Therefore, performance-degradation monitoring alone may fail to identify an environment change, and LCM decisions require more sophisticated mechanisms, such as change detection \cite{VVV2024CD}.
}

%%%%%%%%%%%%%%%%%%%%%%%%%%%%%%%%%%%%%%%%%%%%%%%%%%%%%%%%%%%%%%%%%%%%%%%%%%%%
\section{Two‑Sided Model: Model Management for Interoperability}\label{sec:interop}

Many functions in cellular networks, such as PMI in CSI feedback,  demodulation, and channel decoding, rely on tightly coupled two-sided operations across the UE and gNB. For example, PMI-based CSI feedback works only when both UE and gNB recognize the same mapping from PMIs to precoders. In Section~\ref{sec:lcm}, we have introduced the model management functionalities from the environment and operating scenario change perspectives. When we implement two-sided operations, the model management functional block has to additionally maintain the interoperability between independently developed models. Traditional interoperability is achieved through standardized interfaces. However, for AI-native designs, transmitter and receiver models (e.g., neural encoders and decoders) may be trained separately by different vendors, and these models may have different interfaces and therefore are hard to connect seamlessly with each other, which introduces new technical hurdles for model management, training, and inference.

A representative example from 3GPP is the use case of CSI compression using neural encoders and decoders. In this use case, 
the UE uses a neural encoder to compress the precoder into a latent message and transmits it to the gNB. The gNB then reconstructs the precoder using a neural decoder. While joint training yields high fidelity, it is impractical in multi-vendor deployments.

To enable multi-vendor interoperability, 3GPP is studying training collaboration mechanisms that specify how training-related information is exchanged between UE and gNB vendors. These mechanisms allow each vendor to train its corresponding model independently while preserving compatibility with the counterpart model, i.e., the decoder can correctly reconstruct the intended precoder from the encoder output. This use case highlights that future standardization must address not only signaling interfaces but also training formats, latent representations, and trust mechanisms as ML-based CSI compression becomes part of PHY operations.

\subsection{Collaboration Modes}

In the initial study during Rel-18, 3GPP identified three types of inter-vendor collaborative training \cite{TR38843}:
\begin{itemize}
\item Type 1: Joint training of the two-sided model at a single side/entity, e.g., UE side or network side.
\item Type 2: Joint training of the two-sided model at the network side and UE side, respectively.
\item Type 3: Separate training at the network side and UE side, where the UE-side CSI generation part (encoder) and the network-side CSI reconstruction part (decoder) are trained by the UE side and network side, respectively.
\end{itemize}
Here, ``joint training'' means that the generation and reconstruction models are trained in the same forward-backward loop, either at a single node or across multiple nodes (e.g., via gradient exchange).

%TC:ignore
\begin{figure*}[!t]
  \centering
  \includegraphics[width=0.75\linewidth]{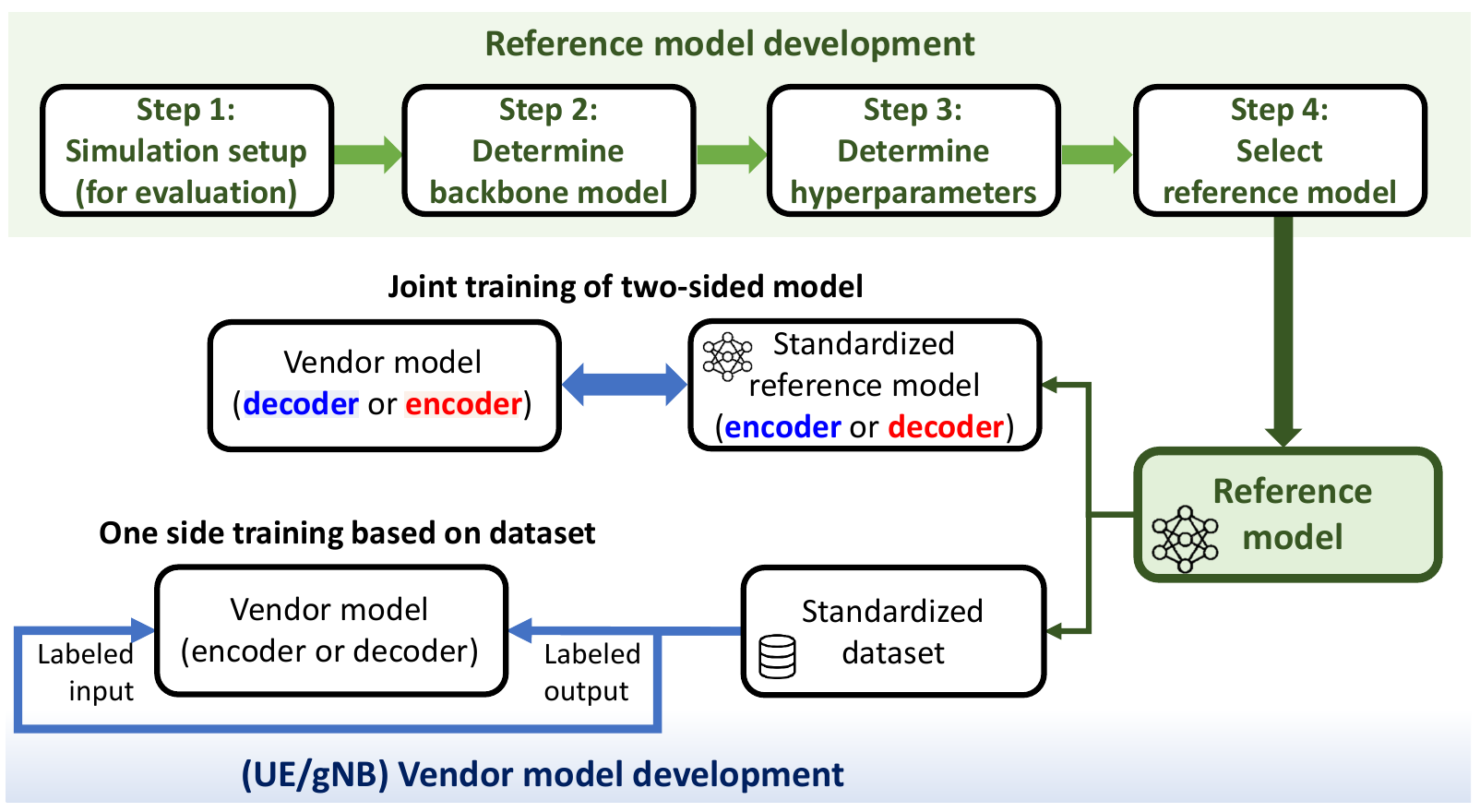}
  \caption{Two-sided model: reference model and vendor model development.}
  \label{fig:twosided}
\end{figure*}
%TC:endignore

The follow-up Rel-19 studies focused mostly on Type 3 inter-vendor collaborative training and its derivatives since Type 1 does not require extensive interface design, and the study on the air interface for Type 2 was deprioritized by 3GPP. In Rel-19, 3GPP concluded the following feasible directions:
\begin{itemize}
    \item Direction I: Providing fully standardized reference model(s) and parameters with specified CSI encoder and/or CSI decoder.
    \item Direction II: Providing the necessary information that enables UE-side offline engineering. The necessary information could be the dataset with \{target CSI, feedback CSI\}, or the encoder parameters together with \{target CSI\}. Target CSI refers to the precoder estimated by the UE and feedback CSI refers to the latent message sent to gNB.
\end{itemize}

For Direction I, a vendor pairs its own model with the standardized reference model and jointly trains it while keeping the reference fixed.  For example, a UE vendor trains its encoder against the fixed reference decoder, and a gNB vendor trains its decoder against the fixed reference encoder. For Direction II, a UE (gNB) vendor uses the \{target CSI, feedback CSI\} dataset as labeled encoder (decoder) input/output to train its own encoder (decoder).
\cite{Korpi-26WCOM} adopts a similar approach in which UE vendors provide \{target CSI, feedback CSI\} datasets to gNB vendors, which train decoders to reconstruct the feedback CSI. 
{In the ongoing Rel-20 WI, both directions are included in the specification discussion. For Direction I, 3GPP considers using the CSI model(s) in the standardized test/verification procedures for UE as the reference model for vendor model development. For Direction II, 3GPP is developing the dataset exchange procedures and formats to enable the exchange across different BS and UE vendors \cite{R12603841}.}
We depict the derivation procedures in Directions I and II in the vendor model development part of Fig. \ref{fig:twosided} {integrated with the proposed reference model development procedure which will be explained in Sec. \ref{sec:ref_dev}}.

Since multiple reference models, parameter sets, and datasets can be specified, the associated ID is introduced to facilitate the pairing of the CSI encoder and CSI decoder based on the collaboration procedure. That is, it is confirmed that the CSI encoder and CSI decoder are collaboratively trained using the same reference model or dataset (and/or parameter set) by checking the associated ID.

\subsection{{Proposed Reference Model Derivation and Selection Procedure}}\label{sec:ref_dev}
Note that to implement Direction I, 3GPP needs to agree on which model(s) to be captured in the specification as the reference model(s). Similarly, for Direction II, 3GPP needs to first conclude the reference model(s) that generate dataset with \{target CSI, feedback CSI\} or the encoder parameters together with \{target CSI\}. Since the reference model(s) can determine how the commercial CSI encoder and decoder are developed, the reference model(s) are important foundations of the commercialization and implementation of the two-sided models in AI-native transceivers. Hence, a carefully designed derivation procedure to develop the reference model(s) for Directions I and II implementation is the first step towards the successful deployment of two-sided models. 

{Given that the derivation procedure of the reference model is one of the most important foundations for successful deployment of the two-sided model, we propose the following procedures to derive the reference model(s) for inter-vendor collaboration (depicted together with vendor model development in Fig. \ref{fig:twosided}):}
\begin{description}
    \item[Step 1]: Determine the simulation setup and encoder input dataset generation procedure for the performance evaluation of the candidate models.
    \item[Step 2]: Determine the model backbone structure, e.g., Convolutional Neural Network (CNN), transformer, or their variants, based on initial assessment of performance, robustness, and complexity. Further detailed structure of the model can be included in this step or Step 4. 
    \item[Step 3]: Determine training hyper-parameters.
    \item[Step 4]: Determine the test decoder reference model(s) based on performance evaluation against the following criteria:
    \begin{itemize}
        \item Achievable performance: It can be intermediate KPIs, e.g., SGCS, or overall system performance of throughput or block error rate.
        \item Complexity: It includes Floating point Operations (FLOPs) or model storage size.
        \item Robustness: Whether the decoder(or encoder) from the reference model can achieve satisfactory performance when connected to encoder(or decoder) with different model backbone structures, but trained using inter-vendor collaborative training procedures.
    \end{itemize}
\end{description}
The inter-vendor collaboration procedure targets the best system performance with reasonable complexity to limit implementation cost, as well as a flexible design space to allow vendors to develop their own models based on the reference model(s). Therefore, the selection of the reference model has to take performance, complexity, and robustness (when connecting to models developed independently by vendors) into consideration.

%%%%%%%%%%%%%%%%%%%%%%%%%%%%%%%%%%%%%%%%%%%%%%%%%%%%%%%%%%%%%%%%%%%%%%%%%%%%
\section{Key Technologies and Future Directions}\label{sec:keytech}
Building on the LCM foundations described above, this section discusses key technical challenges and future directions for AI-native RAN design.
\subsection{Design Challenges in Resource-Efficient Performance Monitoring}
Although 3GPP has laid out a preliminary air-interface design for performance monitoring schemes in beam prediction and CSI feedback use cases, the design of performance monitoring schemes that effectively evaluate AI/ML model performance based on the air-interface specification remains challenging. 

\paragraph*{ i) Resource Allocation Trade-Offs in Monitoring Design}
When scheduling monitoring resources, using more resources enables better tracking of AI/ML model performance but reduces the overhead savings achieved by AI/ML models. For example, in the beam prediction use case, scheduling more monitoring resources, such as transmitting more beams in the to-be-predicted beam set, allows better tracking of AI/ML predictions but diminishes the reference signal overhead reduction benefit provided by the AI/ML models. Therefore, it is a key challenge to design the monitoring resources intelligently to achieve a better trade-off between the monitoring procedure performance and the amount of resources used.

\paragraph*{ii) Monitoring Without Complex Decoder at the UE}
Another key challenge lies in managing complexity. In two-sided CSI compression, the UE uses a lightweight encoder to compress the precoding vector, while the gNB uses a more complex decoder for reconstruction. Since the UE lacks access to the reconstructed precoder, it cannot directly assess downlink quality. Replicating the decoder at the UE would close this gap but would significantly increase device complexity. Efficient monitoring must therefore provide meaningful quality indicators without requiring full decoder functionality at the UE \cite{Guo-25TWireless}.

\subsection{Input Distribution Shifts and Their Impact on Model Reliability}

\paragraph*{i) Detecting Environmental Drift Through Input Statistics}  
When the characteristics or statistics of the data that the AI/ML model operates on deviate from the training data, the performance of the AI/ML model may degrade. In such cases, we rely on the LCM controller to recover the performance. So far, 3GPP has focused mostly on performance monitoring based on the AI/ML model output. In addition to monitoring the model output, the distribution of the model input data can indicate whether the operating environment is starting to deviate from the training dataset of the current AI/ML model. In fact, 3GPP studies on CSI feedback show that changes in the input dataset distribution may degrade AI/ML model performance \cite{202504AIMLWF}. Therefore, tracking variations in model input statistics provides valuable information for LCM controller decisions and can be considered part of the LCM functionality framework as depicted in Fig. \ref{fig:lcm_framework_detailed}. 

\paragraph*{ii) Input Descriptors as Keys for Model Retrieval}  
Furthermore, input statistics play a critical role in selecting appropriate models from storage. The density of models stored in the system, in terms of coverage across the statistical input space, should be sufficient to ensure that adjacent models offer similar performance levels with respect to key KPIs. In this context, input distribution descriptors not only support drift detection but also act as compact and effective retrieval keys for identifying the most suitable model for a given operating condition. {For example, in the CSI case study in Fig. \ref{fig:csi_sim}, the LCM controller decision to correctly select the UMa-based model after the switch may depend on successfully analyzing the characteristics of channel statistics to identify the UMa environment.}

\subsection{Key Challenges in Intelligent LCM Decision-Making}
As depicted in Fig.~\ref{fig:lcm_framework_detailed}, the LCM controller takes performance monitoring and model input statistics collection as input and makes model control decisions. There are several challenges in making accurate model control decisions:

\paragraph*{i) Information Fusion Across Heterogeneous Inputs}
The LCM controller takes model input analysis and performance monitoring as input and leverages assistance signaling from the other side as side information. {How to combine heterogeneous information from different functionalities and collected at different time slots to make accurate and timely model control decisions is crucial for designing an effective LCM controller.}

\paragraph*{ii) Diagnosing Model Misalignment vs. System Limitations}
When system performance degrades, model misalignment may not be the only cause. For example, an SNR drop may lead to degradation in CSI prediction accuracy even when the AI/ML model still captures the channel dynamics properly. Therefore, the LCM controller needs to infer whether the observed performance degradation is due to misalignment between the current AI/ML model and the environment, or whether the system suffers from inevitable performance degradation, such as degradation caused by SNR drops. 

\paragraph*{iii) Selecting Effective Model Control Actions}
% \textbf{Action selection}. 
After deciding that the AI/ML model in use does not fit the current environment, the next challenge for the LCM controller is to select an action from the available model control options, such as switching to another model, adapting the model by adjusting parameters, or falling back to non-AI/ML-based algorithms. The LCM controller needs to derive sufficient information from the controller inputs, including performance monitoring results and model input analysis, to select or adjust the AI/ML model that aligns with the current environment, or fall back to non-AI/ML models if it determines that the current environment cannot be recognized to ensure proper AI/ML model operations. {Moreover, to avoid frequent back-and-forth switching in highly dynamic scenarios, the controller may require RAN-local stabilization mechanisms, such as hysteresis margins or dynamic thresholds, whose design remains an open challenge.}

\subsection{{Rethinking Interoperability vs. Flexibility in AI-native RAN}}
As explained in the previous section, {the 3GPP Rel-19 study concluded that the interoperability of the CSI feedback two-sided model can be guaranteed by standardizing reference model(s) or dataset(s) generated by reference model(s) to ensure interoperability from both training and inference perspectives}. However, standardizing reference model(s) limits the design flexibility of the AI/ML model to adapt to different environments or types of channels in order to produce the best recovered precoders with minimum CSI feedback size. In fact, that flexibility is one of the major advantages of using AI/ML models.

Moreover, standardized reference model(s) may delay the adoption of the latest AI/ML model architectures into the system since the standardized model(s) must be updated before vendors can use new model architectures. Therefore, to fully explore the potential of two-sided AI/ML models in AI-native transceiver design, we need to rethink the approach to guaranteeing the interoperability of two-sided models. Ideally, we need to have a set of specifications that guarantee interoperability when the models on both sides follow the specifications.

%%%%%%%%%%%%%%%%%%%%%%%%%%%%%%%%%%%%%%%%%%%%%%%%%%%%%%%%%%%%%%%%%%%%%%%%%%%%
\section{Conclusion}\label{sec:conclusion}
From the Rel-17 SA5 management framework and the Rel-18 study of AI/ML for the NR air interface to the ongoing Rel-20 work on two-sided CSI compression, 3GPP has progressively developed AI/ML life-cycle support along two related but distinct tracks. The SA5 track provides longer-timescale OAM capabilities for model training, validation, deployment, inference management, and life-cycle governance, while the RAN track addresses time-sensitive model operation, monitoring, applicability assessment, signaling, and fallback over the NR air interface. The ongoing work on inter-vendor collaboration and two-sided model operation provides an important basis for interoperable AI-native transceivers, although several mechanisms remain under development. Further research is needed to balance monitoring overhead, detect distribution shifts, support reliable LCM decisions, and enable flexible model adaptation and training across diverse and evolving RAN environments.

%TC:ignore
{\renewcommand{\baselinestretch}{1.1}
\bibliographystyle{IEEEtran}
\bibliography{References}
%TC:endignore 

\end{document}